\documentclass[prl,aps,twocolumn,showpacs]{revtex4-2}
\usepackage{amsmath,amssymb,graphicx}
\usepackage{textcomp}
\usepackage{color}
\usepackage{hyperref}
\usepackage{bm}
\usepackage[outdir=./]{epstopdf}

\def\rr{\mathbf{r}}

\hypersetup{pdfpagemode=UseNone,colorlinks=true,linktoc=page,pdfborder={0 0 0},linkcolor=blue,citecolor=blue}

\begin{document}

\title{Photon-mediated Stroboscopic Quantum Simulation of a $\mathbb{Z}_{2}$  Lattice Gauge Theory}
\author{Tsafrir Armon}
\address{Racah Institute of Physics, Hebrew University
of Jerusalem, Jerusalem 91904, Israel}
\author{Shachar Ashkenazi}
\address{Racah Institute of Physics, Hebrew University
of Jerusalem, Jerusalem 91904, Israel}
\author{Gerardo Garc\'{i}a-Moreno}
\address{Institute of Fundamental Physics IFF-CSIC, Calle Serrano 113b, 28006 Madrid, Spain}
\address{
Departamento de F\'{\i}sica Te\'orica and IPARCOS, Universidad Complutense de Madrid, 28040 Madrid, Spain}
\address{Instituto de Astrof\'{\i}sica de Andaluc\'{\i}a (IAA-CSIC), Glorieta de la Astronom\'{\i}a, 18008 Granada, Spain}
\author{Alejandro Gonz\'alez-Tudela}
\email{a.gonzalez.tudela@csic.es}
\address{Institute of Fundamental Physics IFF-CSIC, Calle Serrano 113b, 28006 Madrid, Spain}
\author{Erez Zohar}
\email{erez.zohar@mail.huji.ac.il}
\address{Racah Institute of Physics, Hebrew University
of Jerusalem, Jerusalem 91904 , Israel}
\pacs{}

\begin{abstract}
Quantum simulation of lattice gauge theories (LGTs), aiming at tackling non-perturbative particle and condensed matter physics, has recently received a lot of interest and attention, resulting in many theoretical proposals, as well as several experimental implementations. One of the current challenges is to go beyond 1+1 dimensions, where four-body (plaquette) interactions, not contained naturally in quantum simulating devices, appear. In this Letter, we propose a method to obtain them based on a combination of stroboscopic optical atomic control and the non-local photon-mediated interactions appearing in nanophotonic or cavity QED setups. We illustrate the method for a $\mathbb{Z}_{2}$ lattice Gauge theory. We also show how to prepare the ground state and measure Wilson loops using state-of-the-art techniques in atomic physics.
\end{abstract}

\maketitle

Gauge theories are very fundamental in physics. With their local symmetries, they offer a consistent way to mediate interactions between matter particles. In spite of their clear description of fundamental forces, such as those of the standard model, many important phenomena in such models are still hidden behind the nonperturbative curtain. LGTs~\cite{Wilson1974,Kogut1979} have been an important tool for tackling  such problems: combined with Monte-Carlo methods, they have enabled to compute an enormous amount of valuable physical data (e.g.~the hadronic spectrum~\cite{Aoki2020}). While these computations are useful and widely applicable to a variety of physical problems, they cannot deal with real time evolution (since they are performed in a Euclidean spacetime) or important scenarios with finite chemical potential, due to the sign problem~\cite{troyer05a}.

To circumvent these limitations, quantum simulators of LGTs~\cite{wiese_ultracold_2013,zohar_quantum_2016,dalmonte_lattice_2016,banuls_simulating_2020,klco2021standard} based on cold atoms in optical lattices~\cite{Zohar2011,Banerjee2012,zohar12a,Zohar2013,Zohar2013a,zohar_cold-atom_2013,Banerjee2013,Stannigel2013,Bazavov2015,Kuno2015,zohar_digital_2017,zohar_digital_2017-1,Kasper2017,Dutta2017,Gonzalez-Cuadra2017,Bender2018,Zache2018,Rico2018,Ercolessi2018,Magnifico2019}, trapped ions ~\cite{hauke_quantum_2013,yang_analog_2016,Davoudi2020,davoudi2021simulating}, superconducting qubits~\cite{marcos_superconducting_2013,Marcos2014,Mezzacapo2015,Romero2017}, and Rydberg atoms~\cite{Weimer2010,Tagliacozzo2013,tagliacozzo_simulation_2012,Surace2020,Celi2020} have been proposed. These theoretical efforts have already culminated in several experimental realizations~\cite{martinez_real-time_2016,kokail_self_2019,Schweizer2019,mil_scalable_2020,yang_observation_2020}, benchmarking the validity and applicability of the quantum simulation method (as well as quantum computing methods ~\cite{klco_quantum_2018,lamm_general_2019,klco_su_2020,paulson_towards_2020,ciavarella_trailhead_2021,atas2021su2}).
Yet, so far, the experimental implementations have been limited to one-dimensional and/or small systems. One of the current challenges of the field is thus precisely to simulate LGTs in more than $1+1d$. Such simulations are especially hard \cite{Zohar2021} due to the four-body interactions associated with the magnetic plaquette interactions they involve~\cite{kogut_hamiltonian_1975}. The analog proposals to which were based on perturbative constructions, either by enforcing the symmetries~\cite{Zohar2011} or using gauge invariant building blocks~\cite{Zohar2013}, which made the available plaquette interactions very weak. A way to avoid that is to implement the dynamics stroboscopically, such that two-body interactions with ancillary degrees of freedom yield the desired four-body interactions~\cite{weimer_rydberg_2010,Tagliacozzo2013,tagliacozzo_simulation_2012,zohar_digital_2017-1,zohar_digital_2017}. However, despite the many advantages of this approach, if the ancilla-system interaction is local, sequential actions are required for creating the plaquette interactions, each with an interaction of one of the four participating degrees of freedom with the ancilla alone, wasting  precious experimental coherence time. This is the case of \cite{zohar_digital_2017-1,zohar_digital_2017} where the two-body interactions are obtained by adiabatically moving atoms and relying on their collision - a very slow process.

In this Letter, we propose an alternative stroboscopic strategy to obtain such four-body interactions using the non-local, tunable, photon-mediated interactions which appear naturally when placing atoms close to photonic crystals~\cite{shahmoon13a,douglas15a,gonzaleztudela_subwavelength_2015,Hung2016,goban13a,thompson13a,tiecke14a,goban15a,hood16a,Kim2019,Luan2020,Beguin2020,Chang2019,orevic2021} or within cavity mirrors~\cite{ritsch13a,Davis2019,Bentsen2019,Periwal2021}. In particular, by using a judicious combination of magnetic field gradients and time-modulated Raman transitions~\cite{Hung2016,Davis2019,Bentsen2019,Periwal2021}, we will be able to entangle an ancilla array to all the plaquette atoms \emph{simultaneously} and avoid the sequential nature of previous proposals~\cite{zohar_digital_2017-1,zohar_digital_2017}. Such simultaneous action was also proposed in the toric code simulation of~\cite{weimer_rydberg_2010} and the $U(1)$ proposal of~\cite{tagliacozzo_optical_2013} using Rydberg excited states with finite lifetimes. In our case, the physical degrees of freedom are encoded in metastable atomic states, and we use Raman photon-assisted transitions to make them interact. With that tool, we discuss in detail a method to simulate a pure $\mathbb{Z}_{2}$ LGT in $2+1d$. Moreover, we show how to adiabatically find the ground state and measure observables (as Wilson loops), using state-of-the-art AMO techniques. 

\emph{$\mathbb{Z}_{2}$ LGTs.} The interest in $\mathbb{Z}_{N}$ LGTs~\cite{horn_hamiltonian_1979} is twofold. First, their large $N$ limit reproduces compact QED ($U(1)$), and thus they can be used as finite local Hilbert space approximations of it on quantum simulators. Second, $\mathbb{Z}_{N}$ is the center of $SU(N)$, which makes it  relevant for studying $SU(N)$ confinement ~\cite{thooft_on_1978}. We focus 
on the simplest case - $\mathbb{Z}_2$~\cite{wegner_duality_1971}. Quantum simulations of $\mathbb{Z}_N$ LGTs have been proposed~\cite{Zohar2013a,notarnicola_discrete_2015,zohar_digital_2017,Ercolessi2018}, and in particular for the $\mathbb{Z}_2$ case  \cite{zohar_digital_2017,Barbieroe2019,Cui2020a,homeier_Z2_2020,Gonzalez-Cuadra2020,Gustafson2021}. The $1+1d$ case was implemented experimentally in ~\cite{Schweizer2019}.

\begin{figure} [tb]
 \includegraphics[width=\linewidth]{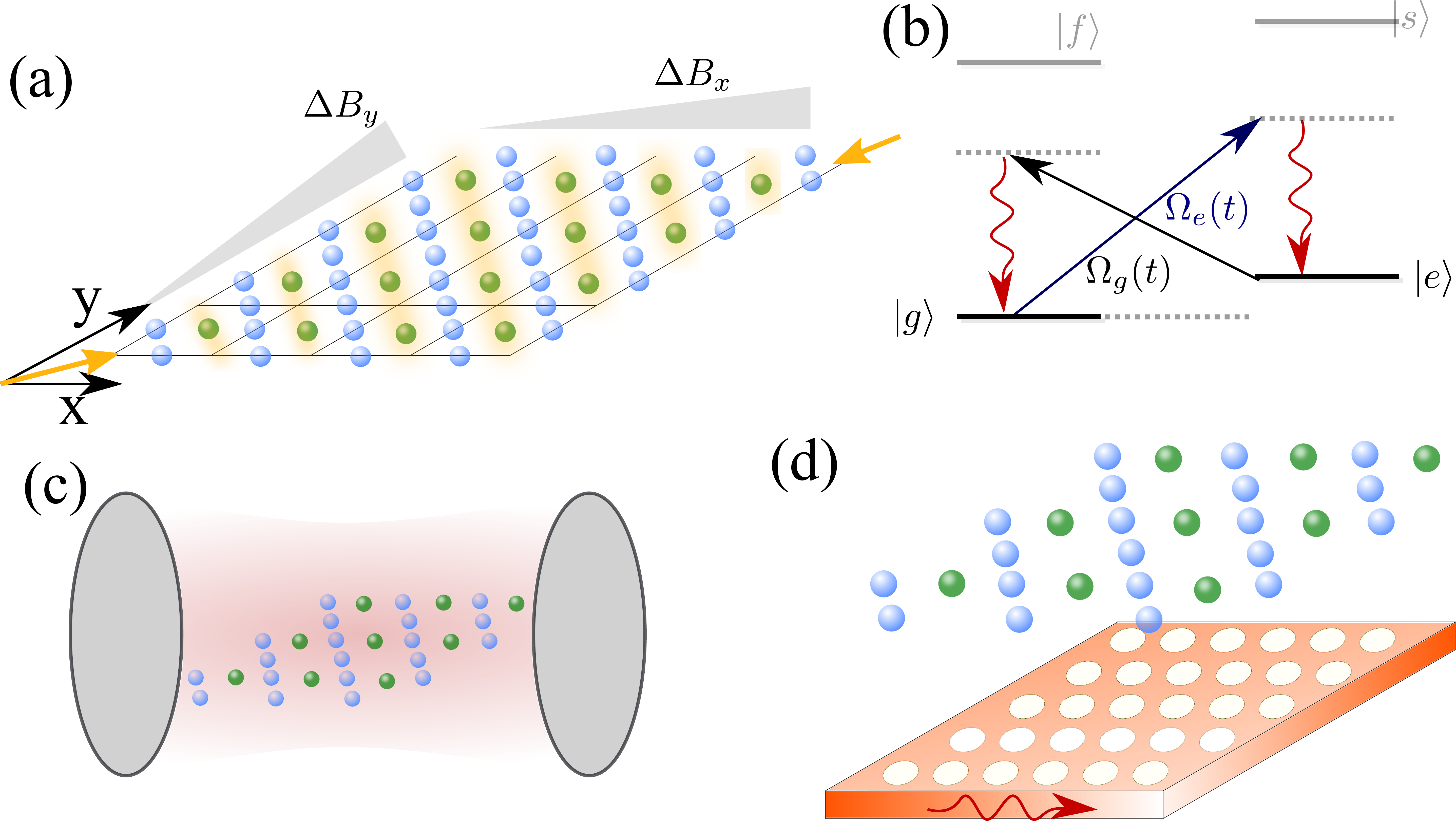}
 \caption{(a) Simulators' scheme: a two-dimensional array of atoms plays the role of the physical degrees of freedom along
  the links (in blue). Another two-dimensional array is interspersed such that one has an additional \emph{control} atoms within the plaquettes (in green). Both the control/link atoms can be controlled independently by changing the relative phase of the lasers generating a diagonal standing wave (in orange). (b) Scheme of the atomic level structure: the $\mathbb{Z}_2$ Hilbert space of the simulator is obtained with two atomic metastable states, $g,e$. We use Raman-photon assisted transitions in a butterfly configuration to induce and control the photon-mediated interactions. The photon field mediating the interactions can be either a cavity field (c) or the guided photons in 2D photonic crystals (d).}
 \label{fig:1}
\end{figure}

 $\mathbb{Z}_{2}$ LGTs are based on the algebra of two anticommuting operators squaring to the identity, on local two dimensional spaces; e.g. the Pauli operators $\sigma_{x}$ and $\sigma_{z}$. We place such a Hilbert space on each link of a two-dimensional lattice (blue spheres in Fig.~\ref{fig:1}(a)), and introduce the local gauge transformations acting on the four links around each site $\mathbf{x}$, 
$A\left(\mathbf{x}\right)=\sigma_{z}\left(\mathbf{x},1\right)\sigma_{z}\left(\mathbf{x},2\right)\sigma_{z}\left(\mathbf{x-\hat{1}},1\right)\sigma_{z}\left(\mathbf{x-\hat{2}},2\right)$,
where the links are labelled by the pair $\mathbf{x},k=1,2$ of their starting site and direction respectively. The dynamics is given by the Hamiltonian $H=H_{E}+H_{B}$~\cite{horn_hamiltonian_1979}, where
$H_{E}=-\lambda_{E}\sum_{\mathbf{x},k}\sigma_{z}\left(\mathbf{x},k\right),$
and
$H_{B}=-\lambda_{B}\sum_{\mathbf{x}}B\left(\mathbf{x}\right)$
are the electric and magnetic parts respectively, and $B\left(\mathbf{x}\right)=\sigma_{x}\left(\mathbf{x},1\right)\sigma_{x}\left(\mathbf{x+\hat{1}},2\right)\sigma_{x}\left(\mathbf{x+\hat{2}},1\right)\sigma_{x}\left(\mathbf{x},2\right)$. Note that $\left[A\left(\mathbf{x}\right),B\left(\mathbf{y}\right)\right]=0$ for any $\mathbf{x},\mathbf{y}$, and thus $\left[H,A\left(\mathbf{x}\right)\right]=0$ for any $\mathbf{x}$, defining a local symmetry. 

\emph{Stroboscopic Evolution.} To mitigate the difficulty of creating the $H_E+H_B$ simultaneously, we turn to a stroboscopic evolution, given by the Trotter formula $e^{-iHt}=\lim_{M\rightarrow\infty}\left(e^{-iH_{E}t/M}e^{-iH_{B}t/M}\right)^{M}$ \cite{Trotter1959a,Suzuki1985a}. The challenge is to implement the unitary operators $e^{-iH_{j}t/M}$, within the simulator, especially  $H_{B}$ with its four-body interactions. We will describe how this stroboscopic evolution was done in~\cite{zohar_digital_2017-1,zohar_digital_2017}, and show how to improve it using tunable photon-mediated interactions. 

The key idea of the method of ~\cite{zohar_digital_2017-1,zohar_digital_2017} is to introduce an array of auxilary degrees of freedom (controls) within the plaquettes (in green in Fig.~\ref{fig:1}(a)), with  the same Hilbert space as on the links (two level systems in our case). The controls (denoted with a $\widetilde{\cdot}$) allow to convert the four-body interaction with a set of four two-body interactions between the control and the links around it. To do it, one creates \emph{stators}~\cite{Reznik2002,Zohar2017b} (see the supplemental material for a brief review) using the unitary controlled operators:
\begin{equation}
\mathcal{U}_{\mathbf{x}}^{(i)}=\left|\tilde{\downarrow}\right>\left<\tilde{\downarrow}\right|_{\mathbf{x}}+\left|\tilde{\uparrow}\right>\left<\tilde{\uparrow}\right|_{\mathbf{x}}\otimes\sigma_{x}^{\left(i\right)}.\label{eq:U}
\end{equation}
acting on the control of a plaquette whose lower left corner is $\mathbf{x}$ and a link $i$ of the same plaquette. The controls are initialized to  $\left|\tilde{\text{in}}\right>_{\mathbf{x}}=\left(\left|\tilde{\uparrow}\right>_{\mathbf{x}}+\left|\tilde{\downarrow}\right>_{\mathbf{x}}\right)/\sqrt{2}$, and the action of $\mathcal{U}_{\mathbf{x}}^{(i)}$ results in the stator $S_{\mathbf{x}}^{(i)}=\mathcal{U}_{\mathbf{x}}^{(i)}\left|\tilde{\mathrm{in}}\right>_{\mathbf{x}}$, mapping states of the physical system (links) to states in the product space of links and controls. $S_{\mathbf{x}}^{(i)}$ obeys  $\tilde{\sigma}_{x}\left(\mathbf{x}\right)S_{\mathbf{x}}^{(i)}=S_{\mathbf{x}}^{(i)}\sigma_{x}^{\left(i\right)}$ - enabling to trade  operations on the links with operations on the controls. Labelling the plaquette links by $i=1...4$ (the order is irrelevant since $\mathbb{Z}_2$ is Abelian), the process could be repeated for all links in the plaquette, to 
\begin{equation}
S_{\mathbf{x}} = \mathcal{U}_{\mathbf{x}}^{(1)}\mathcal{U}_{\mathbf{x}}^{(2)}\mathcal{U}_{\mathbf{x}}^{(3)}\mathcal{U}_{\mathbf{x}}^{(4)}\left|\tilde{\text{in}}\right>_{\mathbf{x}},
\end{equation}
 for which 
\begin{equation}
	\tilde{\sigma}_{x}\left(\mathbf{x}\right)S_{\mathbf{x}}=S_{\mathbf{x}}B\left(\mathbf{x}\right).
\end{equation}
 The  $\mathcal{U}_{\mathbf{x}}^{(i)}$ commute, and thus could also be applied simultaneously, if experimentally allowed (unlike in the previous cold atom proposals \cite{zohar_digital_2017,zohar_digital_2017-1}).
Furthermore, thanks to that all the plaquette interactions can be implemented at once, by acting with the unitary 
$\mathcal{U} = \underset{\mathbf{x}}{\prod}\underset{i\in p\left(\mathbf{x}\right)}{\prod}\mathcal{U}^{(i)}_{\mathbf{x}}$ (where $p\left(\mathbf{x}\right)$ is the plaquette whose lower left corner is $\mathbf{x}$ and $i$ labels its four links, in a counter-clockwise order from $\mathbf{x}$) on an initial state of all the controls,
 $\left|\tilde{\text{in}}\right\rangle = \underset{\mathbf{x}}{\bigotimes}\left|\tilde{\text{in}}\right\rangle_{\mathbf{x}}$.
  To implement $\mathcal{U}$, we require (R1) to have separate control on the links and the controls; and (R2) to allow nearest neighbor control-link interactions of the form $\sigma_{x(z)}\tilde{\sigma}_{x(z)}$.

\emph{Photon-mediated LGT simulator.} The different elements of the simulator are sketched in Fig.~\ref{fig:1}. The basis is a two-dimensional ordered array of ultra-cold atoms as in Fig.~\ref{fig:1}(a) which can be obtained either with optical lattices~\cite{bloch_many-body_2008} or the latest optical tweezer techniques~\cite{endres16a,barredo16a,barredo18a}. Unlike in other proposals, we assume that the atoms do not move, and the interactions among them will be obtained through photon exchange. To control these photon-mediated interactions the atoms are assumed to have a level structure as  in Fig.~\ref{fig:1}(b). The ground states,  $\left|g\right>$ and $\left|e\right>$, will define the needed two level system, whereas the optically excited states $\left|f\right>$ and $\left|s\right>$ will be used for inducing Raman-assisted transitions between the control/link atoms either by coupling them to cavity photons, Fig.~\ref{fig:1}(c), or to the guided modes propagating within a photonic crystal, see Fig.~\ref{fig:1}(d). Individual rotations of each control/link can be obtained with two-photon Raman transitions via the same or another excited state. Let us see in detail how these platforms can potentially fulfil all the required ingredients for the simulators. 

Using two counter-propagating beams with in-plane wave number of $\mathbf{k}_{xy}=\pm\left(\pi/2a\right)\left(\hat{x}+\hat{y}\right)$ (where $a$ is the lattice constant) and phase difference of $\phi$, one can address only the controls (links) by switching $\phi=0$ ($\pi$), as schematically depicted in orange in Fig.~\ref{fig:1}(a). Another more demanding alternative consists of using quantum gas microscopes~\cite{bakr09a,sherson10a,weitenberg11a}, which have already enabled single-site imaging/addressing. In any case, the first requirement (R1) can be satisfied. Let us now go to the non-trivial part of tailoring the two-body interactions between the links and controls. Since the atoms are fixed in our implementation, as compared to~\cite{zohar_digital_2017-1,zohar_digital_2017}, we use the non-local interactions that are induced by the exchange of photons through their optical transition~\cite{Hung2016,Davis2019,Bentsen2019,Periwal2021}. As explained in \cite{douglas15a,Hung2016}, by choosing a \emph{butterfly} Raman configuration as in Fig.~\ref{fig:1}(b), with equal Raman weight $\frac{|\Omega_e|^2}{\Delta_e}=\frac{|\Omega_g|^2}{\Delta_g}\equiv R$, and tuning the effective frequency of the Raman transition far from the cavity frequency or within a band-gap in the cavity/photonic crystal implementations, respectively, one can arrive at the following effective interactions between atoms
\begin{align}
H_{\mathrm{eff}}=J \sum_{j,l}\sigma_{x}^{j}\sigma_{x}^{l}f\left(\rr_{j}-\rr_{l}\right).
\label{HInteraction}
\end{align}
Here, $J$ is the overall spin-spin interaction strength, proportional to $R$, such that the interactions can be switched on and off through the Raman lasers $\Omega_{i}$, and to the atom-cavity/nanophotonic coupling strength. One advantage of the nanophotonic platform is that thanks to the confinement of light, $J$ can potentially be of the order of GHz~\cite{douglas15a,orevic2021}, which could make the simulation considerably faster than other systems. The spatial dependence of the interactions is encoded in $f\left(\rr_{j}-\rr_{l}\right)$, depending on the particular implementation chosen. For cavity QED setups, the interactions have infinite range $|f(\rr_i-\rr_j)|\approx 1$, such that \emph{all} atoms will interact if they are not placed in nodes of the cavity mode. In two-dimensional nanophotonic waveguides~\cite{Gonzalez-Tudela2015b,Hung2016}, the interactions are mediated by an atom-photon bound-state yielding 
\begin{equation}
|f(\rr_i-\rr_j)|\propto e^{-|\rr_{ij}|/L}/\sqrt{|\rr_{ij}|/L},
\end{equation} with a tunable $L$ through the system parameters, such as the Raman laser frequency $\omega_L$. In neither case the interactions take the desired nearest-neighbour form required for the simulation (R2). In the nanophotonic case, one can approximately obtain it by making $L\sim a$. However, the residual couplings to other neighbours break the Gauge symmetry and spoil the simulation (See Sup. Material for details). The way to circumvent this problem was proposed in \cite{Hung2016}, and already experimentally implemented in cavity QED setups~\cite{Davis2019,Bentsen2019,Periwal2021}: it consists of a combination of magnetic field gradients along the X and Y direction, so that each photon-mediated transition at a given distance has a different resonance, plus time-modulated Raman transitions $\Omega_i(t)=\sum_{\alpha} \bar{\Omega}_\alpha e^{i\bar{\omega}_\alpha t}$, where $\bar{\omega}_{a}$ is the detuning from the main laser frequency $\omega_{L}$. Thus, choosing adequately $(\bar{\Omega}_\alpha,\bar{\omega}_\alpha)$ one can design any spatial interaction pattern allowed by $f(\rr_i-\rr_j)$. Concretely, for a magnetic gradient $B=\nabla B\left(p\hat{x}+q\hat{y}\right)$, a Zeeman splitting would be induced at site $m$ with position $(x^m,y^m)$ such that $\omega_{e}^{\left(m\right)}=\mu_{B}\nabla B\left(px^{\left(m\right)}+qy^{\left(m\right)}\right)$. This leads to a different Raman condition for the interaction between each pair of sites that reads $\bar{\omega}_{\alpha}-\omega_{e}^{\left(m\right)}=\bar{\omega}_{\beta}-\omega_{e}^{\left(n\right)}$, which for the desired nearest-neighbors interactions would reduce to $\bar{\omega}_{\alpha}-\bar{\omega}_{\beta}=\mu_{B}\nabla B p$ or $\mu_{B}\nabla B q$, for the horizontal and vertical hoppings, respectively. Here $p,q$ can be, for example, two incommensurable irrational numbers such that the energy difference of vertically or horizontally adjacent sites can not repeat at any other distance. Therefore, only two sidebands $\bar{\omega}_{\alpha}$ would be needed to match the nearest neighbour in the vertical and horizontal directions. The gauge breaking interactions among the different neighbours would then be suppressed as long as the energy differences between the non-nearest neighbouring sites are larger than the overall interaction strength $J$~\cite{Hung2016}. Note, that if such irrational condition can not be satisfied, e.g., due to finite frequency resolution, other larger-distance interactions will be activated, thus limiting the system size that can be simulated without having gauge-breaking terms.

Another potential source of error in the simulation introduced by such photon-mediated interactions is the finite lifetime of the cavity/photonic crystal photons. The effect of these losses in the fidelity of the entangling gates has been well-characterized~\cite{douglas15a}, and scale with $1-F\propto 1/\sqrt{C}$, with $C$ being the cooperativity factor of the system. Recent experiments have already reached values $C\sim 5-30$~\cite{Periwal2021,orevic2021}, although values of $C\sim 10^4$ are feasible with fabrication and trapping improvements~\cite{douglas15a}. Since this error does not decrease with repetition rate, as the Trotter error, there will be a trade-off between cancelling both types of errors that ultimately limit the maximum simulation time. Using a second-order Trotter expansion~\cite{Zohar2017}, this maximum time can be shown to scale $T\propto C^{2/3}$.

\emph{Simulation Step --- }
With the photon-mediated interactions at hand, we can prescribe a single first order trotter step. It is given by
$W_{E}\left(t_{0},\tau\right)W_{B}\left(t_{0},\tau\right)$,
where $t_{0}$ is the the time step's beginning, and $\tau$ is its duration. The electric and magnetic steps are given by $W_{i}=\exp\left[-i\tau H_{i}\left(t_{0}+\tau/2\right)\right]$. The electric term is easily created by a single body operation on the links, $W_{E}\left(t_{0},\tau\right)=V_{z}\left(\tau\lambda_{E}\left(t_{0}+\tau/2\right)\right)$. Here we designate single body operations on the links as $V_{i}\left(\Phi\right)=\exp\left(i\Phi\sum_{j}\sigma_{i}^{j}\right)$ and similarly $\tilde{V}_{i}\left(\Phi\right)$ for the controls. As described above, the magnetic term is realized by creating the stators on every plaquette simultaneously, applying the operation on the controls, and undoing the stators, or
\begin{equation}
W_{B}\left(t_{0},\tau\right)=\mathcal{U}^{\dagger}\tilde{V}_{x}\left(\tau\lambda_{B}\left(t_{0}+\tau/2\right)\right)\mathcal{U},
\end{equation}
 The operator $\mathcal{U}$ relies on single-body operations and two-body nearest neighbors interactions for which one uses $H_{\mathrm{eff}}$ (\eqref{HInteraction}) along with the magnetic field gradient. We discuss the exact form of all these in the supplemental material.
 
 Note that second order trotterization scheme~\cite{trot2nd} could be used without adding applications of $W_{B}$. Because the atom-atom interactions in $W_{B}$ are the main experimental bottleneck (in fidelity terms), these should be favored. Trotterization errors 
 \cite{Heyl2019} of the scheme were studied in \cite{zohar_digital_2017,Bender2018}; importantly, since each step is gauge invariant on its own, gauge symmetry is not violated.

\emph{Adiabatic Evolution --- } We illustrate the dynamics that could be simulated with adiabatic ground state preparation. This should be done with caution, as the model has a phase transition \cite{fradkin_phase_1979} where the gap vanishes and the adiabatic theorem cannot be used \cite{AulettaQM_adiabatic}. We show, however, that it is possible to perform the procedure from both sides of the phase transition, such that one can choose the correct side according to the desired values of $\lambda_{E,B}$, without  crossing the transition. Various numerical and renormalization schemes put the transition slightly over $\lambda_{B}/\lambda_{E}\approx 3$ \cite{crit1,crit2,crit3}.

The straightforward direction is to start with $\lambda_{B}=0$, and set the initial state of the links to the ground state of $H_E$,  $\left|0_E\right>=\underset{\text{links}}{\bigotimes}\left|\uparrow\right>$, which is easy to prepare because it involves only single-atom operations. Then one can adiabatically increase $\lambda_{B}$ over time. The logic remains the same in the second direction, starting with $\lambda_{E}=0$ and initializing the system in the ground state of the magnetic Hamiltonian, $\left|0_B\right>$, which is not as trivial (it is the toric code ground state \cite{kitaev_fault-tolerant_2003}). To do it one can use $\mathcal{U}$  and post-selection projection; as shown in the supplemental material (following \cite{ashkenazi_duality_2021}),
\begin{equation}
\tilde{\Pi}_{x}\mathcal{U}\left|\tilde{\text{in}}\right>\otimes\left|0_E\right>=\left|\tilde{\text{in}}\right>\otimes \left|0_B\right\rangle,
\end{equation}	
where $\tilde{\Pi}_{x}$ projects onto $\left|\tilde{\text{in}}\right>$. It can be realized e.g. by rotating the controls to the $x$ basis and then applying a $\pi/2$ pulse driving $\left|g\right>\leftrightarrow\left|f_{3}\right>$ to another excited state $f_{3}$ which is then allowed to emit to the environment. With the initial ground state in hand, the simulation could proceed by increasing $\lambda_{E}$ over time. The success probability, however, scales exponentially with system size since it post-selects on individual measurements. Alternative approaches to prepare it could be to use engineered dissipative processes~\cite{weimer_rydberg_2010}.

\emph{Measurement --- } One can use the same tools to perform key measurements of the system state. The gauge invariant operators in this case are either $\sigma_z$ operators on the links, or Wilson loops, serving as  confinement probes \cite{Wilson1974}. Here these are products of $\sigma_x$ operators on closed paths. They can be measured within the stator formalism \cite{wilson2} using a single control which gains information on the loop by interacting with each of the links along it as in  (\ref{eq:U}), which can be measured locally. In the supplemental material we show how to do it with our scheme. 

\begin{figure}
\includegraphics[width=\columnwidth]{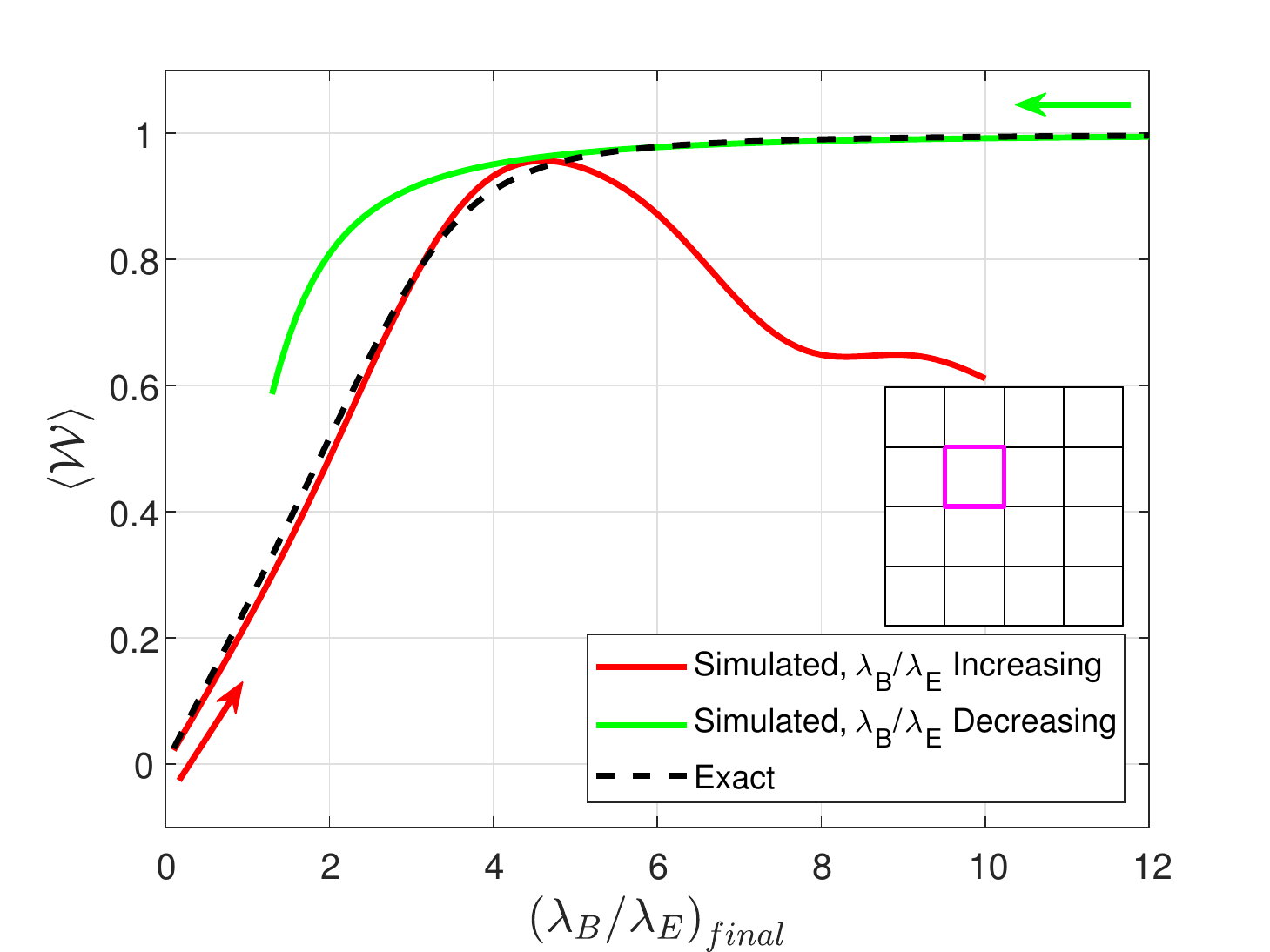}
\caption{$\left<\mathcal{W}\right>$ for  $1\times 1$ Wilson loop in the bulk of a  $4\times 4$ lattice (see inset), as a function of the final ratio $\lambda_{B}/\lambda_{E}$, following an adiabatic preparation of the ground state. The red (green) line shows $\left<\mathcal{W}\right>$ for the adiabatically prepared ground state when $\lambda_{B}/\lambda_{E}$ increases (decreases)  with time (The colored arrows mark the direction of $\lambda_{B}/\lambda_{E}$ change over time). The dashed black line shows $\left<\mathcal{W}\right>$ for the exact ground state. The total simulation time is $1$ with $80$ Trotter steps. }
\label{Fig2}
\end{figure}

\emph{Numerical Simulations --- }
To demonstrate the adiabatic preparation of the ground state, and possible measurement results, we now turn to numerical simulations of a lattice with $4\times 4$ plaquettes. $\lambda_{B}/\lambda_{E}$ is increased/decreased linearly over time to accommodate the two adiabatic processes described above. To keep the simulations comparable (energy scale, frequencies etc.), when starting from the electric ground state the simulated Hamiltonian is $H=H_{E}+\left(\lambda_{B}/\lambda_{E}\right)H_{B}$, while when starting from the magnetic ground state the simulated Hamiltonian is taken as $H=\left(\lambda_{E}/\lambda_{B}\right)H_{E}+H_{B}$. We examine the expectation value $\left<\mathcal{W}\right>$ of a Wilson loop of size $1\times 1$ in the center of lattice, to minimize edge effects, which are dominant in such a small system. Fig. \ref{Fig2} shows the expectation value 
$\left<\mathcal{W}\right>$ for the adiabatically prepared ground states as function of the final ratio $\lambda_{B}/\lambda_{E}$ for the two preparation directions. One can see that the simulations agree with the exact ground state in their respective sides. Due to the small system size, this crossing is not directly related to the phase transition which appears in the thermodynamic limit, but nonetheless demonstrates the advantages of our proposed method since for the same number of trotter steps, better fidelity is achieved by selecting the evolution direction of $\lambda_{B}/\lambda_{E}$, even in the absence of a phase transition (for more details see Supplemental Material).
Even in the thermodynamic limit, for every value of $\lambda_{B}/\lambda_{E}$, except values close to the phase transition, it is possible to adiabatically prepare the ground state by a proper choice of the evolution direction in $\lambda_{B}/\lambda_{E}$. Furthermore, by experimentally measuring $\left<\mathcal{W}\right>$ from both sides, the phase transition itself could be identified by looking for the crossing of the two measurements.

To conclude, we have shown how to use the non-local photon-mediated interactions appearing in cavity and nanophotonic setups to implement a $\mathbb{Z}_2$ LGT simulator. In particular, we demonstrate how the possibility of inducing multi-qubit gates between the ancillary and the physical degrees of freedom allows us to simplify the stroboscopic stator protocol developed in ~\cite{zohar_digital_2017-1,zohar_digital_2017}. 
The stator procedure can be generalized directly to any gauge group~\cite{zohar_digital_2017-1}, but when it is non-Abelian a sequential action is necessary for the plaquettes. The procedure shown here can hence be generalized for studying further 
$\mathbb{Z}_N$ LGTs, also as a method to approximate $U(1)$ physics. It can also be generalized to include  dynamical matter. The ideas introduced in this letter could obviously be adapted in a straightforward manner for the toric code implementation (by adding more controls on the sites, and replacing $H_E$ by $H_A = -\lambda_A \underset{\mathbf{x}}{\sum}A\left(\mathbf{x}\right)$.

\begin{acknowledgements}
\emph{Acknowledgements.} We would like to thank A. Sterdyniak for fruitful discussions.
EZ acknowledges support by the Israel Science Foundation (grant No. 523/20).  A.G.-T. acknowledges financial support from the Proyecto Sinérgico CAM 2020 Y2020/TCS-6545 (NanoQuCo-CM), the CSIC Research Platform on Quantum Technologies PTI-001 and from Spanish project PGC2018-094792-B-100(MCIU/AEI/FEDER, EU). GGM acknowledges financial support from the GEFES prize, State Agency for Research of the Spanish MCIU through the ``Center of Excellence Severo Ochoa'' award to the Instituto de Astrof\'{\i}sica de Andaluc\'{\i}a (SEV-2017-0709), from IPARCOS (Instituto de Física de Partículas y el Cosmos) and from the Spanish Government through the project PID2019-107847RB-C44.
\end{acknowledgements}

\bibliographystyle{apsrev4-2}
\bibliography{RefCombined}

\appendix

\section{Supplemental Material}
\subsection{More on Stators}

Given a composite Hilbert space $\mathcal{H}_A \times \mathcal{H}_B$ describing two physical systems $A$ and $B$,
a stator \cite{reznik_remote_2002,zohar_half_2017} is defined as
\begin{equation}
	S \in \mathcal{O}\left(\mathcal{H}_A\right) \times  \mathcal{H}_B
\end{equation}
where $\mathcal{O}\left(\mathcal{H}_A\right)$ is the set of operators acting on $\mathcal{H}_A$. That makes $S$ a map,
\begin{equation}
	S: \mathcal{H}_A \rightarrow \mathcal{H}_A \times \mathcal{H}_B
\end{equation}
taking a state of system $A$ to a state in the product space $\mathcal{H}_A \times \mathcal{H}_B$. 

In order to construct a stator, we take an "initial" state of system $B$, $\left|\widetilde{\text{in}}\right\rangle_B$, and act on it with a unitary $\mathcal{U} \in \mathcal{O}\left(\mathcal{H}_A \times \mathcal{H}_B\right)$, entangling both systems:
\begin{equation}
	S=\mathcal{U}\left|\widetilde{\text{in}}\right\rangle_B
\end{equation}
For example, consider the case discussed in the main text, where both $\mathcal{H}_A,\mathcal{H}_B$ are two level systems, and we set
\begin{equation}
	\left|\widetilde{\text{in}}\right\rangle = \frac{1}{\sqrt{2}}\left(\left|\uparrow\right\rangle + \left|\downarrow\right\rangle\right)
\end{equation}
and
\begin{equation}
	\mathcal{U}=\left|\uparrow\right\rangle\left\langle\uparrow\right|_B+
	\left|\downarrow\right\rangle\left\langle\downarrow\right|_B \otimes \sigma_{x,A}.
\end{equation}
These give rise to the stator
\begin{equation}
	S=\frac{1}{\sqrt{2}}\left(\mathbf{1}_A \otimes \left|\uparrow\right\rangle_B + \sigma_{x,A} \otimes \left|\downarrow\right\rangle_B\right)
\end{equation}

Stators can be built (by properly choosing the state $\left|\widetilde{\text{in}}\right\rangle$ and the unitary $\mathcal{U}$ \cite{zohar_half_2017}) to satisfy an \emph{eigenoperator} relation,
\begin{equation}
	O_B S = S O_A
\end{equation}
for a pair of operators $O_A \in \mathcal{O}\left(\mathcal{H}_A\right)$ and  $O_B \in \mathcal{O}\left(\mathcal{H}_B\right)$., allowing us to convert operations on $A$ to operations on $B$ and vice versa, in a systematic way. 
In our example, we have $O_A=O_B=\sigma_x$.

If the operators are hermitian and $f\left(x\right)$ is some analytical function of $x$, we also have  
\begin{equation}
	f\left(O_B\right) S = S f\left(O_A\right)
\end{equation}

Thus we can think of a scenario in which we would like to evolve some \emph{physical} system $A$ in time with $H_A$, for example - to implement a single Trotter step, which is hard to implement using our experimental apparatus. Instead, suppose we are able to implement a unitary of the form $\mathcal{U}$ which creates a stator with a control degree of freedom, $B$, initialized in the appropriate state $\left|\widetilde{\text{in}}\right\rangle_B$, giving us a stator satisfying
\begin{equation}
	H_B S = S H_A
\end{equation}
Then, if we can implement $H_B$ evolution, we are done; whatever the initial state of the physical system $\left|\psi\right\rangle_A$ is, we have:
\begin{widetext}
	\begin{equation}
		\mathcal{U}^{\dagger}e^{-iH_Bt}\mathcal{U}\left|\psi\right\rangle_A \otimes \left|\widetilde{\text{in}}\right\rangle_B
		=\mathcal{U}^{\dagger}e^{-iH_Bt}S\left|\psi\right\rangle_A
		=\mathcal{U}^{\dagger}Se^{-iH_At}\left|\psi\right\rangle_A = \left(e^{-iH_At}\left|\psi\right\rangle_A\right) \otimes\left|\widetilde{\text{in}}\right\rangle_B
	\end{equation}
\end{widetext}
That is, we prepare the right state of the control degree of freedom; we act on its product with the arbitrary state of our physical system with some entangling unitary $\mathcal{U}$ creating the stator; we perform a local operation on the control, and then undo the stator - which brings us back to a product of the initial state of the control with the physical state, this time evolved in time using $H_A$.

This can be very useful when one needs to implement complicated many-body interactions, such as the plaquettes in lattice gauge theories \cite{zohar_digital_2017,zohar_digital_2017-1}. As explained in these references, using the stators, one can implement the four-link interaction of the physical system using the unitary $\mathcal{U}$ which only includes two-body link-control interactions, and the local operation $e^{-iH_Bt}$ on the control. This is the method used in the scheme proposed in the main text.

Fur further reading, we refer the reader to reference \cite{reznik_remote_2002}, where stators were introduced, reference \cite{zohar_half_2017}, where their mathematical description was extended and generalized having many-body interactions in mind, and reference \cite{zohar_digital_2017,zohar_digital_2017-1}, where they were utilized for lattice gauge theories. In this work we focus on simplifying the stator quantum simulation methods of lattice gauge theories, as explained in the main text.

\subsection{Creating the simulation step}
\label{AppA}
The simulating system allows one to create single site operations on the links or the controls which we designate by $V_{i}\left(\phi\right)=\exp\left(-i\phi\sum_{j}\sigma_{i}^{j}\right)$ where $i=x,y,z$ and the summation is over all links, or all controls for $\tilde{V}_{i}\left(\phi\right)$. This allows for the creation of the electric Hamiltonian and the electric portion of the time step via $W_{E}\left(t_{0},\tau\right)=V_{z}\left(-\tau\lambda_{E}\left[t_{0}\right]\right)$.

To create the magnetic Hamiltonian and the magnetic portion of the time step, one needs to create the stator by applying 
$\mathcal{U}=\prod_{\mathbf{x}}\prod_{i\in p\left(\mathbf{x}\right)}\mathcal{U}^{(i)}_{\mathbf{x}}$ where the first (second) product is over all plaquettes (links around the plaquette). $\mathcal{U}^{(i)}_{\mathbf{x}}$ is given by $\exp\left(-i\frac{\pi}{4}\tilde{\sigma}_{z}\left(\mathbf{x}\right)\sigma_{x}^{\left(i\right)}\right)\exp\left(-i\frac{\pi}{4}\sigma_{x}^{\left(i\right)}\right)\exp\left(-i\frac{\pi}{4}\tilde{\sigma}_{z}\left(\mathbf{x}\right)\right)$ but because many of the operators commute and could be carried out simultaneously, in terms of the control operations available in the simulating system, $\mathcal{U}$ is simply carried out as
\begin{equation}\label{U}
	\mathcal{U}=\left[\tilde{V}_{y}\left(\frac{\pi}{4}\right)\right]^{\dagger}V_{I}\tilde{V}_{y}\left(\frac{\pi}{4}\right)V_{x}\left(\frac{\pi}{4}\right).
\end{equation}
Here $V_{I}=\exp\left(-i\frac{\pi}{4}\sum_{\left<i,j\right>}\sigma_{x}^{i}\tilde{\sigma}_{x}^{j}\right)$ where the summation is over all nearest-neighbors control-link pairs. $V_{I}$ is created by applying Hamiltonian (2) of the main text for the appropriate duration. The unitaries $\tilde{V_{y}}$ in Eq.~\eqref{U} serve to transform $\tilde{\sigma}_{x}$ in $V_{I}$ to $\tilde{\sigma_{z}}$ to allow for the creation of the stator.

Then a single trotter step $W_{E}\left(t_{0},\tau\right)W_{B}\left(t_{0},\tau\right)$ could be expressed as follows:
\begin{equation}
	\begin{split}
		W_{E}\left(t_{0},\tau\right)&=V_{z}\left(-\tau\lambda_{E}\left[t_{0}\right]\right),\\
		W_{B}\left(t_{0},\tau\right)&=\mathcal{U}^{\dagger}\tilde{V}_{x}\left(-\tau\lambda_{B}\left[t_{0}\right]\right)\mathcal{U},
	\end{split}
\end{equation}

\subsection{Undesired Interactions}
\label{AppE}
Our proposal calls for two-body nearest neighbors interaction, via the use of a tailored magnetic gradient and driving field. It is nonetheless interesting to examine the effects of undesired interactions of atoms further apart from each other. Supoose we include in the interaction term $V_{I}$ other pairs of atoms than the nearest neighbors, the added terms could be split into three categories - link-link interactions, control-control interactions and control-link interactions.

Undesired link-link interactions in $V_{I}$ will have the form $\sigma_{x}^{i}\sigma_{x}^{j}$, and will thus commute with every other term in $W_{B}$ (which for the links only involves $\sigma_{x}$). Therefore, within a single time step, the application of $\mathcal{U}$ and $\mathcal{U}^{\dagger}$ will cancel out the effects of all such undesired interactions. Undesired control-control interactions, however, do not commute with $W_{B}$, but they do cancel out almost entirely for consecutive time steps. If we designate $\mathcal{U}_{\mathrm{actual}}=\mathcal{U}\mathcal{U_{\mathrm{CC}}}$ where $\mathcal{U}_{\mathrm{CC}}$ is the contribution of the undesired control-control interactions, such that $\left[\mathcal{U},\mathcal{U}_{\mathrm{CC}}\right]=0$, $\mathcal{U}_{\mathrm{CC}}$ involves only the controls and thus commutes with $W_{E}$. Therefore, a single time step could be written as:
\begin{equation}
	\mathcal{U}_{\mathrm{CC}}^{\dagger}W_{E}\left(t_{0},\tau\right)W_{B}\left(t_{0},\tau\right)\mathcal{U}_{\mathrm{CC}},
\end{equation}
and therefore, following consecutive time steps $\mathcal{U}_{\mathrm{CC}}$ will cancel out except for $U_{\mathrm{CC}}$ ($\mathcal{U}_{\mathrm{CC}}^{\dagger}$) in the first (last time step). Thus, it will be possible to cancel out all control-control undesired interactions with two applications of two-body interactions targeted only at the controls. This could be done in a similar fashion to the proposed single-body operations only on the controls or the links.

The last group includes undesired control-link interactions. This group is the most problematic one, as it doesn't commute with $W_{B}$ or $W_{E}$, and can not be readily canceled out. Furthermore, these interactions break the gauge symmetry. It should be noted that in the context of photonic waveguide with its spatially declining coupling, the undesired interactions are exponentially suppressed, with the shortest undesired control-link interaction $\sqrt{5}$ times larger than the desired interaction distance. If this suppression is not sufficient, one can split the lattice into sublattices which will be addressed sequentially, such that atoms with undesired couplings will be even further apart.

\subsection{Adiabatic evolution in a (small) finite size system}
\label{AppB}
The $\mathbb{Z}_{2}$ LGT exhibits a phase transition in the thermodynamic limit ($N\rightarrow\infty$ for system of size $N\times N$ plaquettes), but naturally, any experiment will be carried out with a finite size system, and our simulations are limited to small systems (up to $4\times 4$ plaquettes). While the energy gap for finite systems does not vanish like in the thermodynamic limit, the gap decreases with system size and also depends on the ratio $\lambda_{B}/\lambda_{E}$. Therefore, even though small systems do not exhibit the phase transition, the decreasing energy gap recreates some of the difficulties posed by the phase transition, due to the adiabaticity demands. The smaller the energy gap, the longer the process must be to obey the adiabaticity condition \cite{AulettaQM_adiabatic}, and more trotter steps will be needed. Thus, even for a finite (and even small) system it is hard (though not impossible) to adiabatically prepare the ground state for every value $\lambda_{B}/\lambda_{E}$ when starting from either side of the $\lambda_{B}/\lambda_{E}$ axis. Even though our numerical simulations do not exhibit the expected phase transition of the thermodynamic limit, they still demonstrate how choosing the evolution direction in $\lambda_{B}/\lambda_{E}$ can improve the fidelity of the process, without extending the process duration and adding more trotter steps.

\begin{figure}
	\includegraphics[width=\columnwidth]{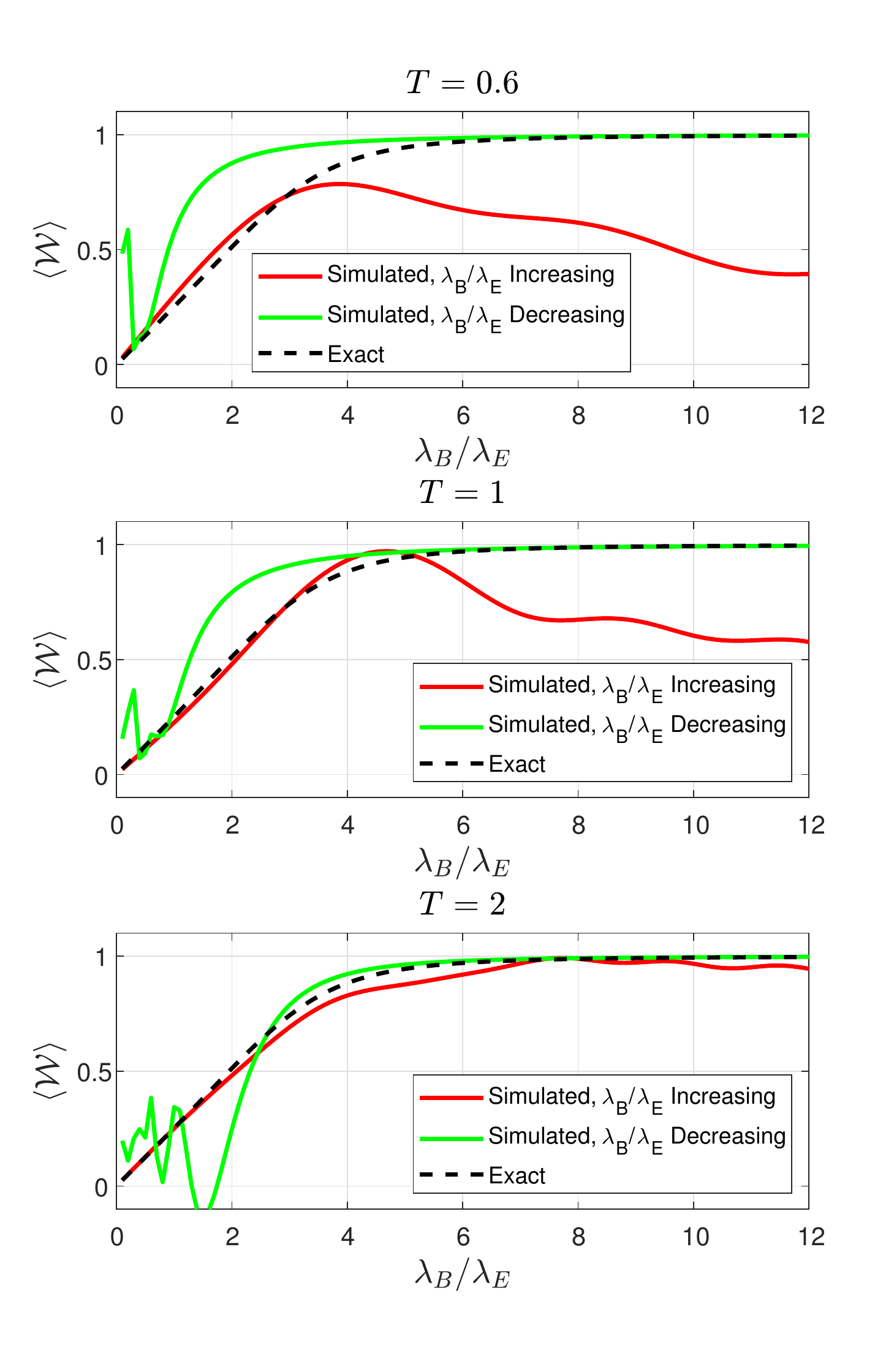}
	\caption{Similar to Fig. 2 of the main text but for for a $1\times 1$ Wilson loop in the center of a $3\times 3$ plaquettes lattice. The duration $T$ of the process is $0.6,1,2$ for the top, middle and bottom panels respectively. The trotter step duration is kept the same so the number of trotter steps $M$ scales with $T$ and is $48,80,160$ in the same order.}
	\label{FigSup1}
\end{figure}

Figure (\ref{FigSup1}) shows similar results to those of Fig. 2 of the main text for a $1\times 1$ Wilson loop in the center of a $3\times 3$ plaquettes lattice. However, the duration of the process is increased from the top panel to the bottom one. One can observe that as the duration of the process increases the two directions of preparation better agree with the exact result - as a result of the increased adiabaticity. However, by choosing the appropriate direction one can reach the desired fidelity with shorter duration (and less trotter steps). This effect will not be possible in the thermodynamic limit where the vanishing energy gap will not allow for adiabatic preparation beyond the phase transition (no matter the duration of the process).

\subsection{Initializing $H_{B}$ ground state}
\label{AppC}
One can easily verify that the ground state of $H_{B}$ is given by
\begin{equation}
	\left|0_B\right>=\prod_{\mathbf{x}}\left(\frac{1+\prod_{i\in p\left(\mathbf{x}\right)}\sigma_{x}^i}{2}\right)\left|0_E\right>,
\end{equation}
where the outer product is over all plaquettes, the inner product is over all links in that plaquette and $\left|0_E\right>=\underset{\text{links}}{\bigotimes}\left|\uparrow\right>$ is the ground state of $H_E$. To initialize the ground state, we note that 
\begin{equation}
	\mathcal{U}\left|\tilde{in}\right>\otimes\left|0_E\right>=\prod_{\mathbf{x}}\left(\frac{1}{\sqrt{2}}\left|\tilde{\uparrow}\right>_{\mathbf{x}}+\frac{1}{\sqrt{2}}\left|\tilde{\downarrow}\right>_{\mathbf{x}}\otimes\prod_{i\in \mathbf{x}}\sigma_{x}^{i}\right)\left|0_E\right>.
\end{equation}
In terms of the eigenvectors along the $x$ direction the term in the product reads
\begin{equation}
	\left|\tilde{\uparrow_{x}}\right>_{\mathbf{x}}\otimes
	\frac{1+\prod_{i\in \mathbf{x}}\sigma_{x}^i}{2}
	+\left|\tilde{\downarrow_{x}}\right>_{\mathbf{x}}\otimes
	\frac{1-\prod_{i\in \mathbf{x}}\sigma_{x}^i}{2},
\end{equation}
and therefore 
\begin{equation}
	\tilde{\Pi}_{x}\mathcal{U}\left|\tilde{in}\right>\otimes\left|0_E\right>=\left|\tilde{in}\right>\otimes \left|0_B\right\rangle
\end{equation}
where $\tilde{\Pi}_{x}$ is a product of projection operators onto the "up" eigenstate of $\tilde{\sigma}_{x}$.

\section{Measurement of Wilson Loops}
\label{AppD}
To measure a Wilson Loop, that is,
\begin{equation}
	\mathcal{W}\left(\mathcal{C}\right) = \underset{\ell \in \mathcal{C}}{\prod} \sigma_x\left(\ell\right)	
\end{equation}
where $\mathcal{C}$ is some closed path on the lattice, and $\ell$ sums over the links along it, one must apply controlled unitaries to entangle the relevant control with each link along the loop, creating a stator \cite{wilson2}. 

\begin{figure} [tb]
	\includegraphics[width=\columnwidth]{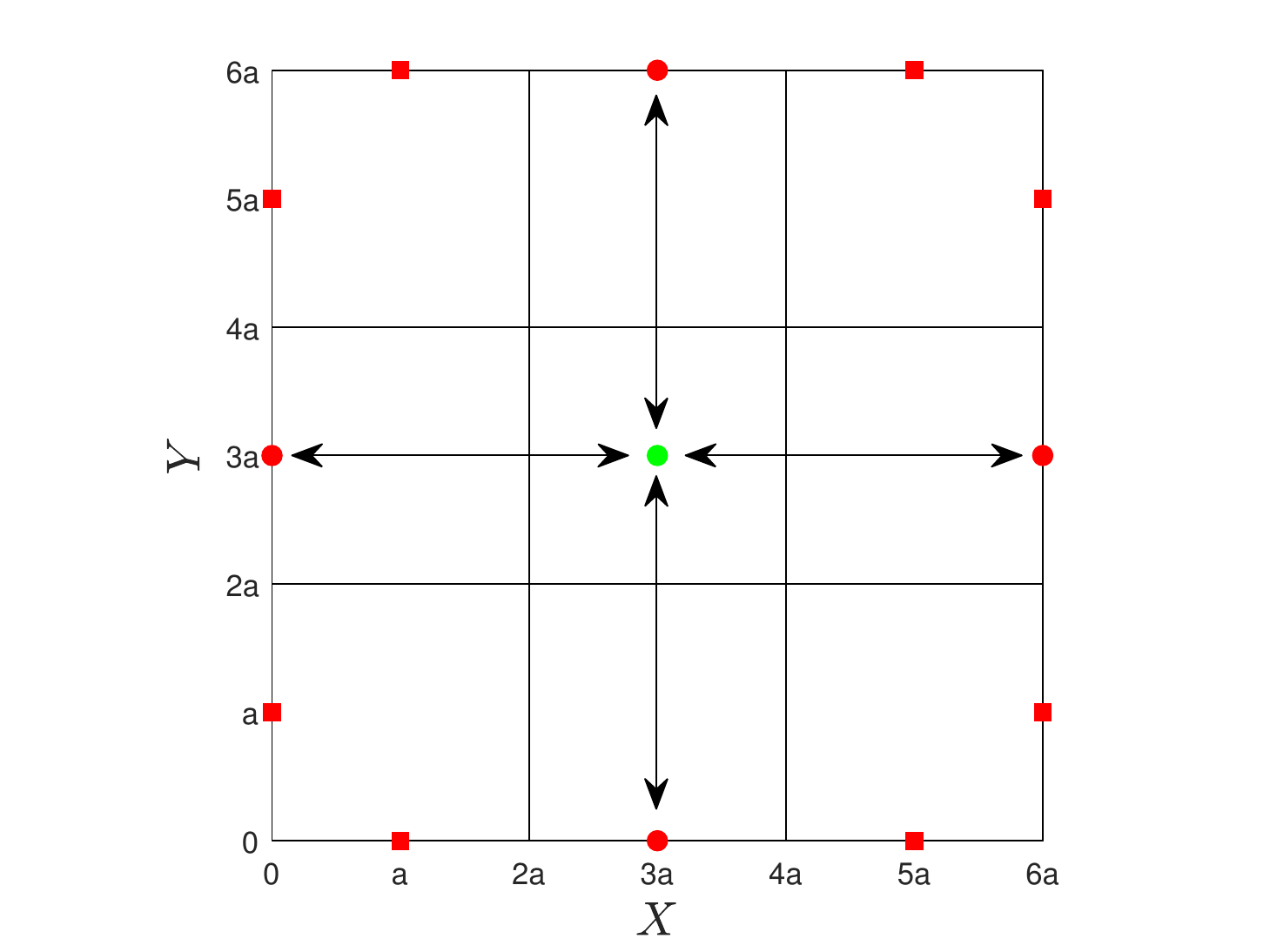}
	\caption{The links (red) and control (green) involved in a $3\times 3$ Wilson loop (other controls and links are omitted). The links are split into two sets of equal distance from the control (circles vs. squares). The arrows demonstrate the simultaneous application of the unitary on the first set.}
	\label{FigSup2}
\end{figure}

Unlike in previous proposals where this process takes place sequentially, with the simulator discussed here, many steps could be performed simultaneously. 
First, consider square loops of size $N \times N$, with $N$ odd. We examine the loop and a control which is located at or near its center. Then, we divide the links into sets of identical distances from the control. For choice of a loop, the control lies exactly at the center, thus minimizing the number of sets (this number scales linearly with $N$). For each set, the magnetic gradient and driving field must be adjusted to select its distance as the desired interaction distance (in a similar fashion to the choice of nearest-neighbor interactions before). It should be noted that in the case of photonic waveguides the interaction length $L$ is tuneable and even long range interactions are feasible~\cite{gonzaleztudela_subwavelength_2015,douglas15a}. An operator $\mathcal{W}$, analogous to $\mathcal{U}$ introduced above but connecting only the  control and links relevant for the Wilson loop should then be applied, creating  the desired a stator. This will generally involve repeating the process for every set of links (with different distances from the control) until we  yield the desired total stator, and the results could be measured on the control. Figure~\ref{FigSup2} presents the links (red) and the control (green) in a $3\times 3$ Wilson loop. The links are split to only two sets of equal distance from the control (squares vs. circles in the figure). The stator creation could take place using only two steps - applying the procedure on all the links of one set (say the circles as exemplified by arrows in the figure) and then on all the links in the other set.

Seeking for Wilson loops in other shapes - note that these can be composed out of smaller square loops with odd sizes, $\mathcal{W}_i$. Note that thanks to the fact that $\sigma_x^2$ is the identity, $\mathcal{W} = \underset{i}{\prod} \mathcal{W}_i$, for any subdivision to smaller loops covering each plaquette enclosed within $\mathcal{W}$ exactly once. This way one can optimize the process of computing the expectation value of a Wilson loop with a minimal number of steps. Also note, that since the important property for probing confinement is the decay law of the logarithm of $\left\langle \mathcal{W} \right\rangle$ for large rectangular loops in the thermodynamic limit, and whether it is proportional to the area of the loop (confined static charges) or its perimeter (deconfined static charges) \cite{wilson_confinement_1974}, for large systems and loops, square loops of an odd size would be more than efficient, because of the very different scaling of $N^2$ vs $4N$ when $N \gg 4$.

It should be noted that we are not limited to the measurement of a single Wilson loop. In fact, applying the steps presented here will create the stators for every Wilson loop of the desired size - simultaneously - each one encoded on the control which lies at its center. Thus, measurements could be taken more efficiently than in previous proposals.

\end{document}